# Behaviour of Ion Acoustic Soliton in two-electron temperature plasmas of Multi-pole line cusp Plasma Device (MPD)

Zubin Shaikh[1,3], A. D. Patel[1,4], P. K. Chattopadhyay[1,2], Joydeep Ghosh[1,2], H. H. Joshi[3] and N. Ramasubramanian[1,2]

1. Institute for Plasma Research, Gandhinagar-382428, India
2. Homi Bhabha National Institute, Anushaktinagar-400094, Mumbai, India
3. Department of Physics, Saurashtra University, Rajkot-360005, India
4. Government Science College, Gandhinagar, Gujarat-382016 ,India

zubin.ipr@gmail.com

## ABSTRACT

This article presents the experimental observations and characterization of Ion Acoustic Soliton (IAS) in a unique Multi-pole line cusp Plasma Device (MPD) device in which the magnitude of the pole-cusp magnetic field can be varied. And by varying the magnitude of the pole-cusp magnetic field, the proportions of two-electron-temperature components in the filament-produced plasmas of MPD can be varied. The solitons are experimentally characterized by measuring their amplitude-width relation and Mach numbers. The nature of the solitons is further established by making two counter-propagating solitons interact with each other. Later, the effect of the two-temperature electron population on soliton amplitude and width is studied by varying the magnitude of the pole cusp-magnetic field. It has been observed that different proportions of two-electron-temperature significantly influence the propagation of IAS. The amplitude of the soliton has been found to be following inversely with the effective electron temperature ($T_{eff}$).

## 1. Introduction

Ion-Acoustic Solitons (IAS), the self-organized, non-linear localized structures that can propagate long distances, are widely studied in astrophysical and laboratory plasmas to understand their non-linear dynamics describing several fundamental processes of plasma physics. A solitary wave was first discovered to propagate long distances without changing its shape and speed in shallow water in 1844 by J.S.Russels[1]. Korteweg-de Vries[2] (KdV) developed the mathematical formalism of soliton dynamics, which had led to significant advances in determining the properties of solitons. The soliton dynamics for plasmas was explained by Washimi and Taniuti[3] by deriving and solving the KdV equation in plasma medium using a novel mathematical approach. Sagdeev[4] studied the arbitrary amplitude non-linear waves in plasma, highlighting many more interesting phenomena of plasma acoustic modes. Both these methods are used to analyze solitons in laboratory[5] and space plasmas[6]. The theory of solitons was further developed by Gardner[7], obtaining an exact solution of the KdV equation by treating it as an initial value problem using the inverse scattering method. T.Taniuti[8] introduced the reductive perturbation principle for solving the KdV equation, which was also applicable for solving various non-linear equations apart from waves in plasma. The solution to the KdV equation for non-linear ion-acoustic waves in plasma medium comprising of negative ions had been obtained by Das and Tagare[9] and Das[10]. A comprehensive review on theoretical studies of solitons can be found in the references[11,12].

The existence of solitary waves is ubiquitous in space plasmas. Solitary waves in the magnetosphere were first observed by the S3-3 satellite[13] and subsequently confirmed and studied extensively by Viking Satellite[14]. These solitary structures were also observed in the auroral acceleration region[15] and the generation mechanism of these structures has been given by Q.Lu[16].

The existence of ion-acoustic solitons in laboratory plasmas was experimentally demonstrated by H. Ikezi[5] for the first time in a double plasma device[17]. In this experiment, the solitons are excited by applying perturbation into a fine mesh grid immersed in plasma with different waveforms having different frequencies[18]. Following this, ion-acoustic solitons were excited in several devices[19–22] mainly by pulsing a floating wire grid placed inside the plasma.

The experiments in the laboratory plasmas demonstrated the variations in soliton properties with varying plasma parameters, which were successfully modelled using the theoretical formulations mentioned above. The experimentally measured width and propagation velocity of solitons matched very well with the solitary wave solutions, as shown by Sakanaka[23].

The soliton-soliton interaction was also experimentally demonstrated by exciting two solitons and making them propagate towards each other[5,24,25]. Over the years, several experiments were carried out in order to understand the excitation[26,27], propagation[28,29] collisions[30], etc., of ion-acoustic solitons in different plasma devices.

Although the ion-acoustic solitons in plasmas are extensively studied both theoretically and experimentally, several open questions remain to be answered, such as the behaviour of solitons in plasmas with two electron temperatures. The coexistence of two distinct species of electrons at different temperatures is very common in space plasmas[13,31] and laboratory plasmas[32,33].

Theoretical studies by Cairns[34] and Nishihara[35] had shown that the presence of non-thermal electrons or plasmas with two electron temperatures can significantly modify the ion acoustic solitary structures. The propagation of ion-acoustic waves in two-electron temperature plasma has been studied by Jones et al.,[36] both experimentally and theoretically, and it has been shown that a small fraction of hot electrons can affect the propagation of the wave. However, there are very few controlled experimental studies on soliton characterization with respect to soliton width and amplitude in plasmas with two-temperature electrons.

In the present work, the effect of two-electron temperature on the properties of ion-acoustic solitons are studied in the Multi-pole line cusp Plasma Device (MPD)[32,33]. The special feature of this device is the controllability of the magnitude of the pole magnetic fields by varying the currents in electromagnets used for producing the cusp magnetic fields. This controllability of cusp magnetic field strength at the poles facilitates the production of two-temperature electrons in variable proportions. The confinement of the high energetic electron population generated in the filament-produced argon plasmas varies with varying the cusp magnetic field strengths leading to the generation of plasmas with two-electron temperatures in variable proportions in this device. Taking advantage of this unique feature, the effect of two-temperature electrons on ion-acoustic soliton propagation and characteristics are studied in the MPD. The solitons are excited using the conventional technique by pulsing a floating metal mesh grid placed inside the plasma. Before varying the magnetic field strengths, the solitons are thoroughly characterized by measuring the velocities and width of solitons and compared with theoretical estimations. The solitary nature of the excited waves is also verified by inducing interactions of two counter-propagating solitons. The cusp magnetic field strengths are then varied to vary the hot-electron fraction. It has been observed that the soliton amplitude increases, and its width decreases with systematically increasing the hot-electron populations in the plasma. Most importantly, it has been found that the effective electron temperature $(T_{eff})$ controls the width of the solitons almost proportionally.

The paper is organized as follows. Section 2 describes the details of the experimental setup and wave excitation and detection techniques. Section 3 describes the detailed experimental results and characterization of solitons. Section 4 describes the effect of two-temperature electrons on soliton dynamics, followed by a summary in section 5.

## 2. Experimental Setup

The present experiment is carried out in a Multi-pole line cusp Plasma Device (MPD)[32,33]. MPD consists of six rectangular-shaped electromagnets with profiled core material to produce the variable multi-pole line cusp magnetic field. These electromagnets provide uniqueness in varying the magnetic field strength and configuration. These electromagnets are placed on the periphery of the vacuum vessel, and each magnet is placed 60 degrees apart. Most early experiments used permanent magnets to create the multi-cusp field. Those devices were limited in terms of performing investigations with variable cusp magnetic fields as the magnetic field produced by permanent magnets is fixed. The novel aspect of MPD is that by varying the current in the electromagnets both in terms of magnitude and direction, the magnetic field strength and configuration can be changed in a controlled way. In all configurations, the magnetic field strength in the center always remains very small in the order of a few gausses only. Hence, the field-free region inside the chamber does not change appreciably. The present experiment is performed with 12 pole cusp magnetic field configuration. The current in all six electromagnets is in the same direction; hence all six magnets will produce one type of pole, and another virtual pole will be produced in between two magnets; hence a total of 12 cusps will be there of six dipoles. For 12-cusp configuration, the magnetic field value remains less than ~ 2 Gauss up to R ~ 8 cm (diameter ~ 16 cm)[37]. We consider this region (up to R ~ 8 cm) as field-free.

The MPD comprises of a non-magnetic stainless steel cylindrical vacuum vessel with a length of 1500 mm, a diameter of 400 mm, and a wall thickness of 6mm. The chamber is evacuated by a Turbo Molecular Pump (440 l/s) backed by a rotary pump through a conical reducer at one side of the chamber. A base pressure of $1 \times 10^{-6}$ mbar is achieved, measured by a hot ionization gauge. The filamentary argon discharge plasma is produced using a hot filament-based cathode source. The plasma source (cathode) is two dimensional (8cm x 8cm) vertical array of five tungsten filaments; each filament has a 0.5 mm diameter and 8 cm length. These filaments are powered by a 500 A, 15 V floating power supply, usually operated at around 16 - 19 A and 7.5V per filament. The chamber is filled with Argon gas through a needle valve to a working pressure of $\sim 8 \times 10^{-5}$ m-bar. The filament source is biased negatively at a voltage of 50 V with respect to the grounded chamber walls using a discharge power supply of ratings

~125V and ~25A. The electric field lines between the high-potential filament and the grounded chamber wall accelerates the primary electrons emitted from the joule-heated filaments. These electrons collide with the neutral argon atoms and ionize some of them. These ions and electrons are confined by the multi-pole line cusp magnetic field.

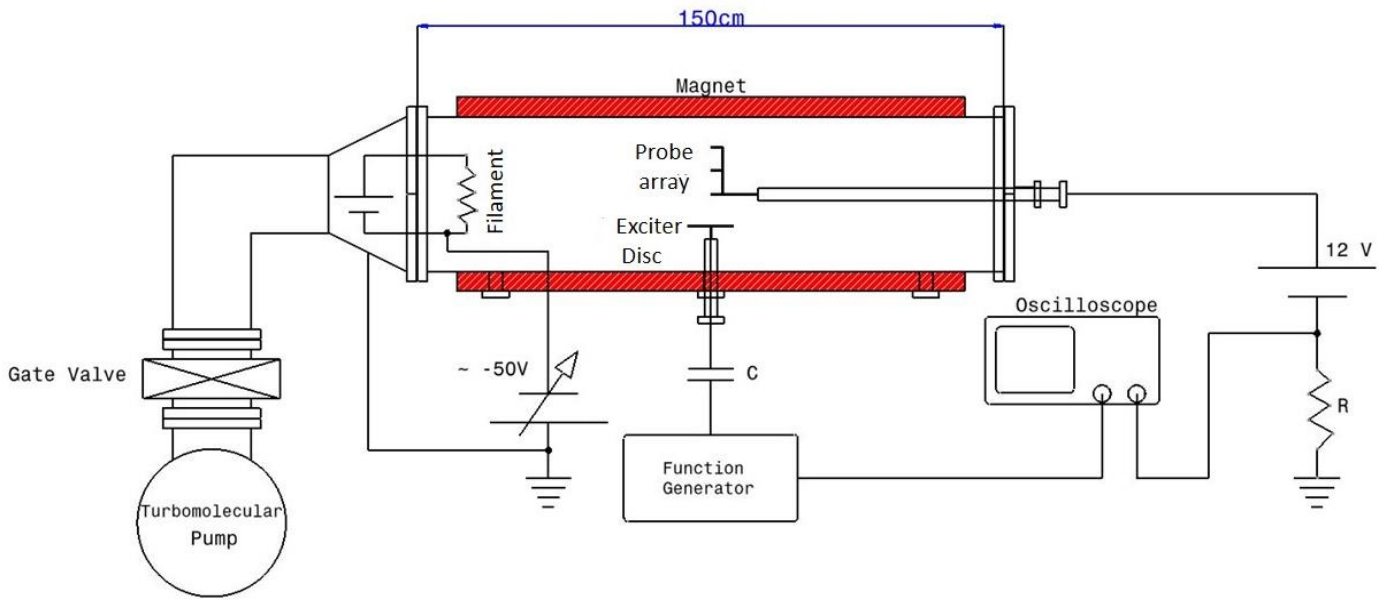

*Figure 1: Schematic diagram of the Multi-Cusp Plasma Device (MPD) Experimental setup*

The plasma thus confined in the MPD is usually reported as bi-Maxwellian, i.e., having two temperature electrons[32,38–40]; one having low temperature called cold electron ($T_c$) and other having relatively larger temperature called hot electrons ($T_h$). It is well known that the cusp magnetic is an ideal configuration for the confinement of primary or hot electrons [41–44]. The cusp magnetic field confines the hot electrons by mirror effects. Due to the mirror effect in cusp configuration, electrons will move back and forth between the poles; thus, the maximum hot energetic electrons are confined by the cusp magnetic field[41,43,45].

The measurement of plasma parameters have been estimated from the V-I characteristics i.e., the variation of probe-current (I) with respect to the variation in probe-voltage (V) applied to the Langmuir probes[46–48]. Although obtaining the V-I characteristics is relatively uncomplicated, extracting plasma parameters from it is not straight forward and requires rigorous and careful analysis. The Langmuir Probe (LP) data, i.e., the V- I curve, obtained from multiple sweeping cycles of the probe-voltage is smoothed first to carefully remove digital and high frequency noise without any loss of information. The smoothing procedures includes fitting a straight line in the sampled ramp voltage and generating the voltage data point from the equation of the fitted straight line. The acquired probe current is then smoothed using the Savitzky-Golay technique[49]. The floating potential ($V_f$) is obtained when the probe draws zero current, $I_p = 0$. The plasma potential ($V_p$) is determined from the maximum of the first derivative of the probe current. Next, the ion current ($I_{ion}$), which varies weakly with the probe potential, is subtracted from the probe current ($I_p$) by using the dependence of the square of the probe current on probe potential. A straight line is fitted in the ion saturation region of the V-I characteristics. The fitted straight line is extrapolated to the plasma potential to obtain the total ion current ($I_{ion}$) [50–52]. The fitted curve is then subtracted from the total probe current ($I_p$) to obtain the electron current ($I_e$). The total current and the total electron current after ion-current removal is shown in figure 2[53–55]. The variation in probe current with respect to the variation in probe voltage is shown in figure 2 by green line whereas the electron current, i.e., after subtracting the ion-current is shown with the red line. The first derivative of the electron current is shown with the black curve.

The $ln\, I_e$ vs $V_{pr}$ is plotted in figure 3. The figure clearly shows two distinct linear regions indicating presence of two-temperature electrons. The hot electron temperature ($T_h$) is estimated from the second slope of $lnI_e$ vs $V_{pr}$ plot as shown in figure 3. A straight line (green color line) is fitted to the second slope (Point B in figure 3) to determine the hot electron temperature ($T_h$). The fitted straight line is extended to a vertical line drawn from the plasma potential ($V_p$) to the voltage axis (shown by dotted line in figure 3). The intersection of the fitted straight line to the second slope and the vertical line from plasma potential ($V_p$) (point D in figure 3) gives the hot-electron current ($I_h$) [56]. After subtracting the hot electron current ($I_h$) from the total current ($I_e$) the cold-electron current ($I_c$) is determined. A straight line (red color line) is fitted to the first slope (point A in figure 3) to determine cold electron temperature ($T_c$) [57].

In bi-Maxwellian plasma, the electron current collected by a probe in the electron retarding region is given by

$$I_e = I_{c0}exp\left[\frac{e(V_{pr} - V_p)}{k_B T_c}\right] + I_{h0}exp\left[\frac{e(V_{pr} - V_p)}{k_B T_h}\right] \qquad (1)$$

Where, $I_{c0} = eA_p N_c(k_b T_c/2\pi m_e)^{1/2}$ and $I_{h0} = eA_p N_h(k_b T_h/2\pi m_e)^{1/2}$, $N_c$ and $N_h$ are the cold electron density and hot electron density, $A_p$ is probe area, and $V_{pr}$ is probe potential, $V_p$ is plasma potential. $N_0 \approx N_c + N_h$ is the total electron density. The hot electron density ($N_h$) is defined as

$$N_h = 4 I_h / eA_p \sqrt{8 T_h e / \pi m_e} \qquad (2)$$

where, $m_e$ is the mass of an electron, $A_p$ is probe area, $T_h$ is temperature of hot electrons, $I_h$ is hot electron component, $e$ is electron charge.

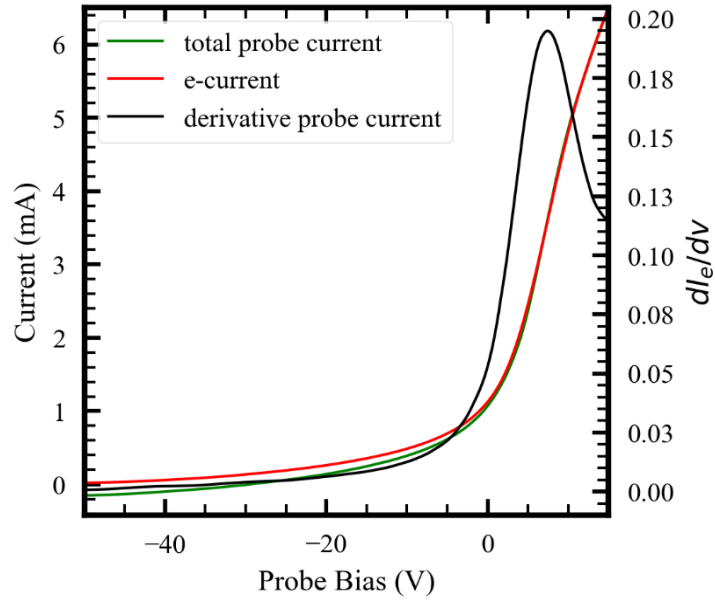

*Figure 2: Variation of probe current to probe voltage and variation of the first derivative of the probe current with probe voltage*

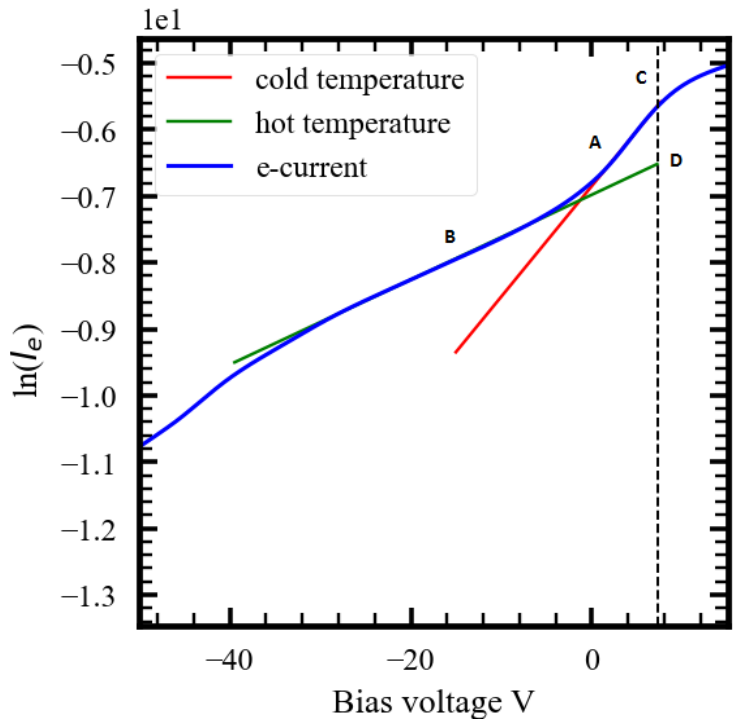

*Figure 3:* Variation of $lnI_e$ with bias voltage

Previously, we have reported that a change of the currents in the electromagnets changes the pole cusp magnetic field, which in turn changes the leak width and mirror ratio[33]. The leak width shows the opposite variation to the cusp magnetic field, i.e., as the magnetic field increases, leak width decreases[33,58,59]. As a result, the high energetic electron confinement in plasma increases, and it affects the plasma parameters. So as expected, the pole cusp magnetic field controls the density and temperature of the plasma. Hence, we varied the $T_c$, $T_h$, $N_c$ and $N_h$ by systematically varying the pole magnetic field, $B_p$. Once $T_c$, $T_h$, $N_c$ and $N_h$ are measured experimentally, effective plasma temperature ($T_{eff}$) can be estimated, which is defined as[36,60]

$$T_{eff} = N_o T_h T_c / (N_h T_c + N_c T_h) \qquad (3)$$

Figure 4 shows the variation of $N_c$ and $N_h$, and figure 5 shows the variation of $T_c$ and $T_h$ at the center of the device (R=0 cm) with different pole cusp magnetic field ($B_p$) strength for argon plasma produced with a fill pressure of ~8 x 10$^{-5}$ mbar. As $B_p$ increases, the leak width[61] $d = 2(r_{le} r_{li})^{1/2}$ changes from 11 mm for $B_p \sim 0.16\ kG$ to 6 mm for $B_p \sim 0.3\ kG$, where the pole separation is 2πr/6 = 20 cm. As the ratio of pole width and pole separation decreases, the plasma confinement increases leading to an increase in the density of both cold and hot ($N_c$ and $N_h$) electrons.

Figure 5 shows that the temperature of cold and hot electron ($T_c$ and $T_h$) initially falls up to $B_p \sim 0.3\ kG$ and then start increasing again. The exact nature of the variation of $N_e$ and $T_e$ (particularly $T_e$) needs a detailed device simulation. However, the aim of the present experiment is to study the effect of $T_{eff}$ on the amplitude and width of the soliton. Figure 6 shows the variation of $T_{eff}$ with $B_p$. $T_{eff}$ is calculated using the measured values of $T_c, T_h, Nc,$ and $N_h$. It has been observed that the $T_{eff}$ decreases initially with increasing the $B_p$ and then starts increasing again.

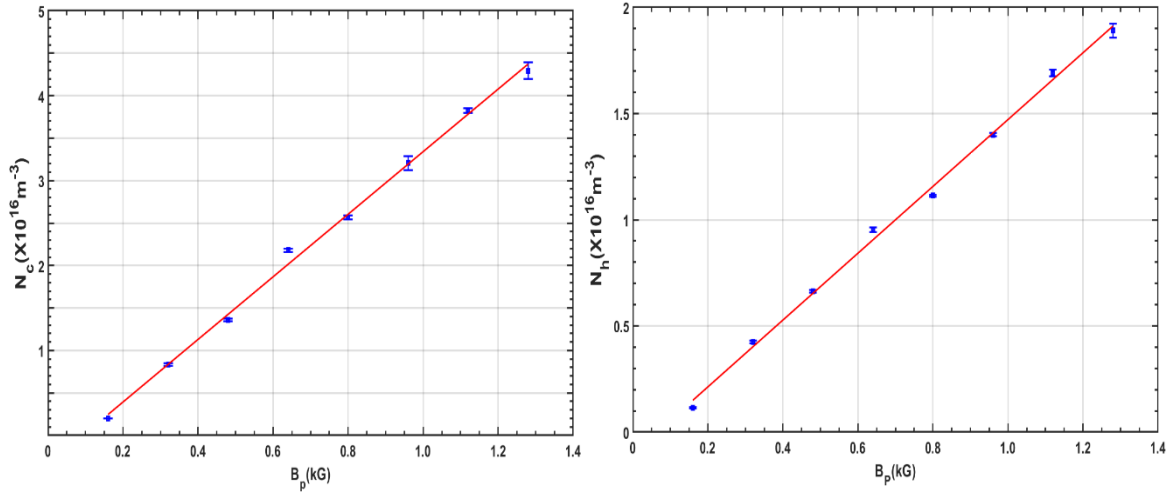

*Figure 4: Variation of cold electron density $N_c$ and hot electron density $N_h$ with pole-cusp magnetic field strengths.*

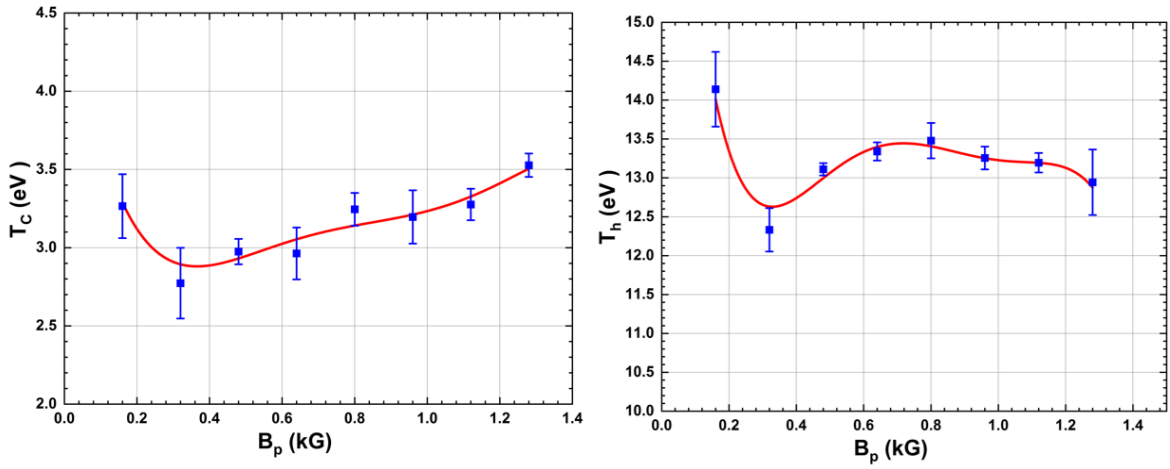

*Figure 5: Variation of cold electron temperature ($T_c$) and Hot electron temperature ($T_h$) with pole-cusp magnetic field strength*

All the basic plasma parameters, such as electron temperature ($T_e$), plasma density ($n_e$), plasma potential ($V_p$), floating potential ($V_f$), and density fluctuations $\delta I_{isat}/I_{isat}$ remains almost constant over the radial extent of the plasma column[32,33,37]. Typical measured plasma parameters at the midplane of the device are: Plasma Density ($n_e$) ~ $10^{16}$ m$^{-3}$, electron temperature ($T_e$) ~ 4-5 eV, Plasma potential ($V_p$) ~ 4-5 V, Electron Neutral Collision frequency ($\nu_{en}$) ~6 x $10^6$ Hz and Ion plasma frequency ($\omega_{pi}$) ~ 4 x$10^6$ Hz for -50V discharge voltage and 8 x$10^{-5}$ bar working pressure and at a pole magnetic field of $B_p = 0.6 kG$ (Magnet Current $I_{mag} = 80A$). Pole Magnetic field ($B_p$) is the measured magnetic field at the pole surface of electro magnets.

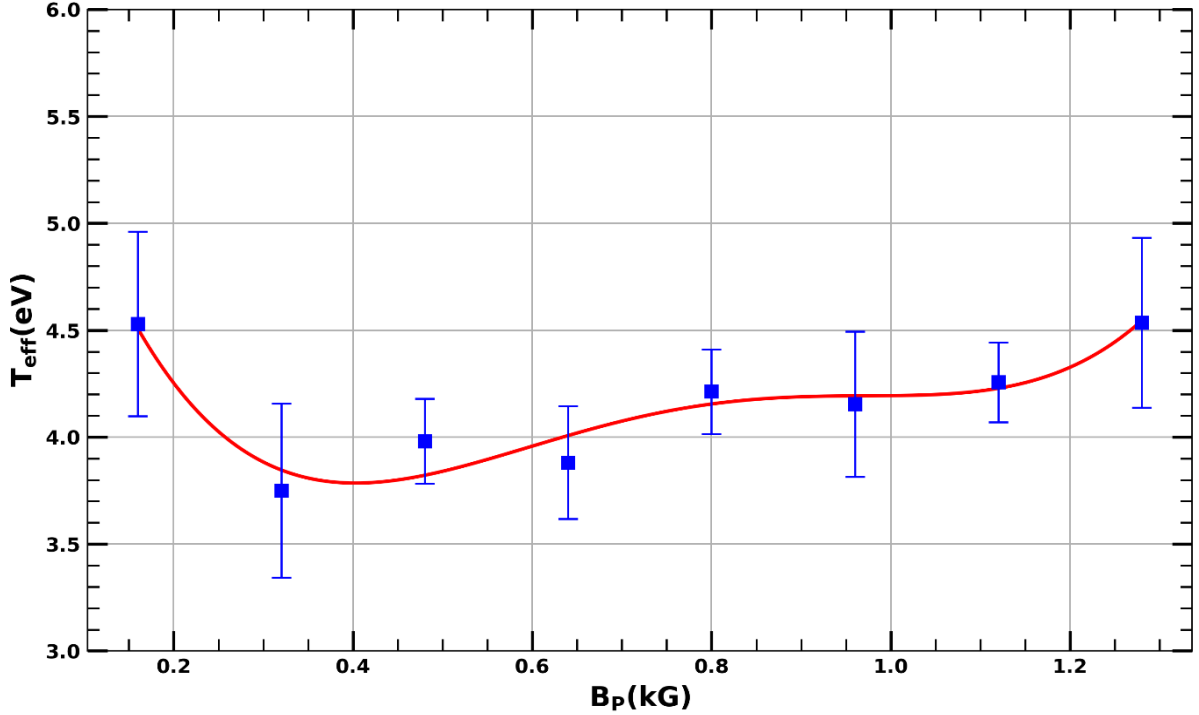

*Figure 6: Variation of effective temperature with pole-cusp magnetic field strengths.*

## 2.1 Wave excitation and detection techniques

After measuring plasma density and temperature, experiments on wave excitation are carried out. Soliton is excited by a floating exciter metal disk which is placed inside the plasma at the central plane of the device, where the magnetic field is minimum, and the plasma is uniform and quiescent[37]. This exciter grid is a solid disk of molybdenum with a diameter of 50 mm and a thickness of 0.25 mm. The solid disk has been used to generate uniform sheath thickness around it[62,25]. A single pulse sinusoidal voltage of ~ 20 $V_{pp}$ at 90 $kH_z$ frequency has been applied to the grid to excite the soliton in the MPD plasma. The magnitude of the voltage perturbation is chosen to be much higher than plasma potential ($V_p$) for soliton excitation, and the perturbation frequency satisfies $\omega/\omega_{pi} < 0.7$[18]. A PA-85 amplifier-based circuit with a gain of ten has been used to amplify the perturbation signal of 90 $kH_z$ generated by a function generator.

A compressive pulse and a rarefactive pulse that follows the compressive pulse are excited to perturb the plasma[62,63]. Plasma's response to this perturbation is detected in the electron saturation regime ($\delta I_{esat}$), at three different radial locations using a set of Langmuir Probes (LP) placed at 2 cm, 6 cm, and 10 cm, respectively, from the exciter disk. All three probes are in a uniform quiescence plasma region where all the plasma parameters are constant, and the magnetic field is minimum[37,64].

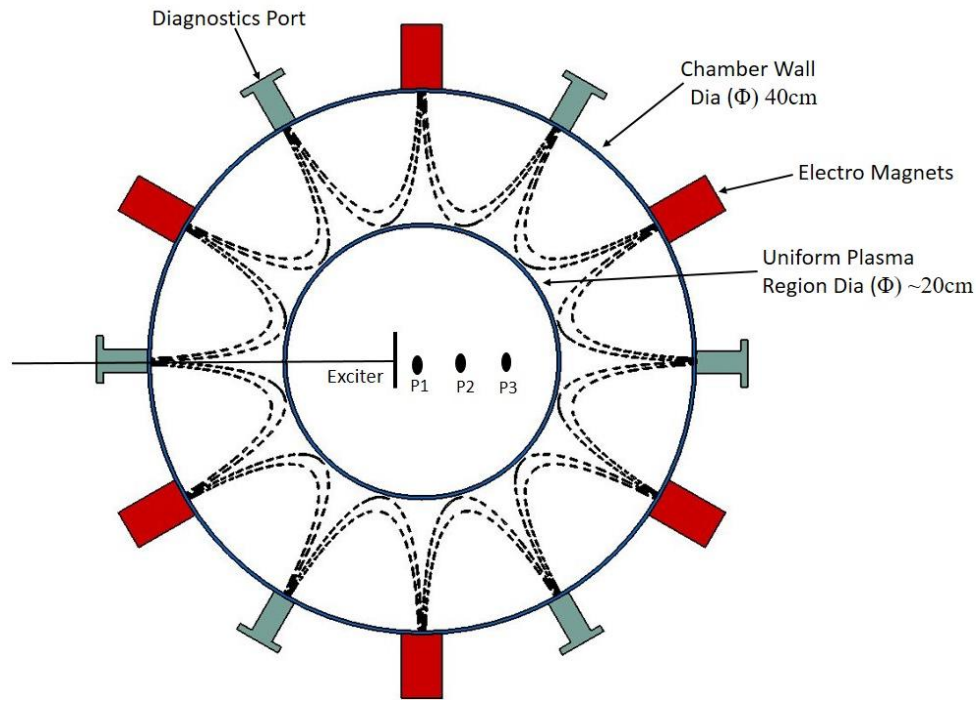

*Figure 7: Cross-sectional view of the experimental setup of the variable multi-cusp magnetic field plasma device (MPD)*

Figure 7 shows the cross-section view of the device with the exciter and Langmuir probe array location. The perturbation is launched in the radial-vertical plane of the device, and a response is also recorded along the same plane. Each LP has a probe-tip diameter of 1 mm, and a tip length of ~5 mm. The probes are biased close to the local plasma potential ~12 V in order to measure the electron saturation current ($\delta I_{esat}$). The data are acquired using an 8-bit digital oscilloscope with different sampling rates and stored for further analysis.

## 3. Soliton wave excitation and its characterization

In our experiment, the soliton has been excited by sinusoidal pulse[5,18] as described below. The time traces of sinusoidal perturbation signal and the plasma's response to applied perturbation is shown in figure 8. The perturbation trace is shown on the top of the figure, and the bottom three traces show the detector probe signals from different LPs placed at different radial locations. These signals are normalized with their maxima, as shown in the figure; hence the Y-axis of all subfigures is on a -1 to 1 scale. Following the application of the pulse to the grid, two distinct pulses separated in time are recorded by the LPs. These two pulses are due to different phenomena occurring after the application of the pulse to the grid, as described below.

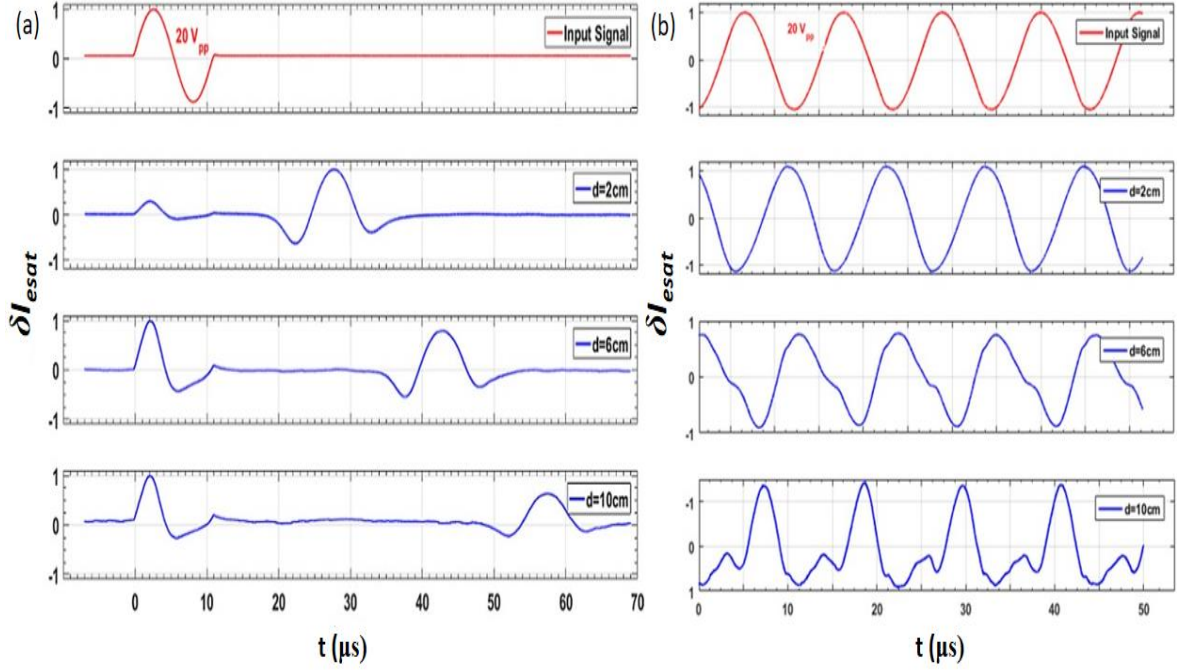

*Figure 8: (a) Time evolution of single pulse sine wave perturbation recorded at the different radial locations; (b): time evolution of continuous sinusoidal wave perturbation recorded at different radial locations*

A near replica of the applied pulse to the grid appears on the LPs almost simultaneously with the applied pulse. This is a signature of an Ion burst or Ballistic mode[65–67]. Although this signal is interpreted as fast electron response, capacitive pickup, or noise. The root cause of its appearance is due to free streaming of ions[66,67]. However, an elaborate explanation of this behaviour is not the main focus of this study. After recording the ion bursts almost simultaneously with grid applied pulse, the time traces of all the LPs shows another pulse with a time delay of ~20 μs. This second pulse in LPs is somewhat a mirror image of the applied pulse. The reason is simple, as plasma responses to the sinusoidal perturbation in opposite polarity, a negative peak followed by a positive peak is detected by LPs. Note that the grid perturbation has a positive peak followed by a negative peak. The shape of time-delayed pulses acquired by the LPs does not remain same as that of applied perturbation. The negative half of the time trace recorded by the LPs does not follow a sine variation and shows a shape relevant to the hyperbolic secant. From the observed time delay in the received signal, the velocity of the wave has been calculated by the time of flight method. The width of the wave and the velocity are compared as a function of its amplitude. To obtain this relation, the amplitude of the grid perturbations pulse is varied from 10 $V_{pp}$ to 20 $V_{pp}$.

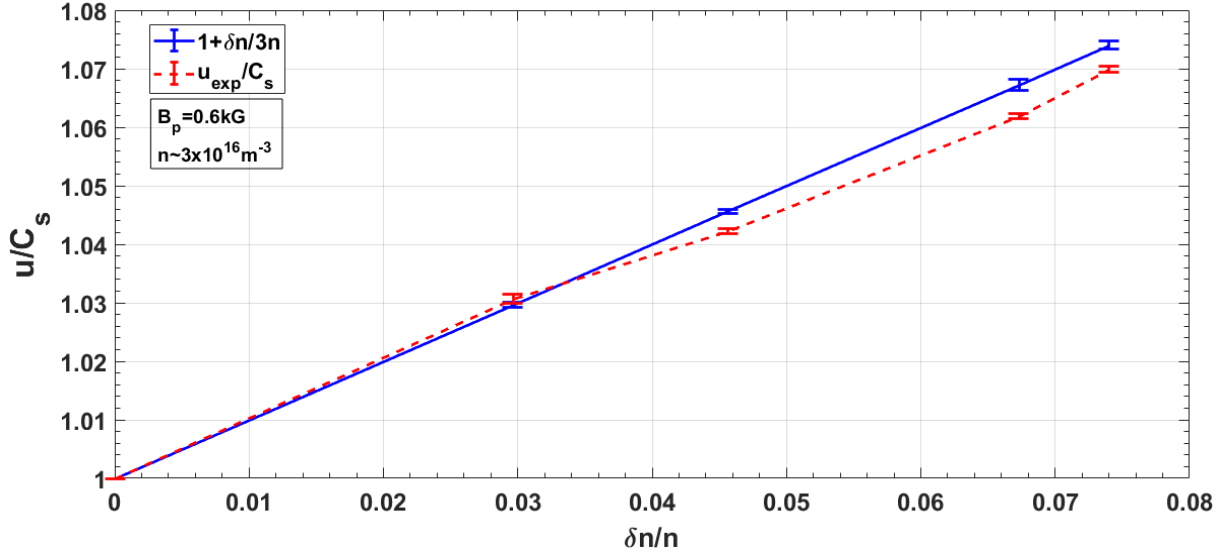

*Figure 9: Velocity of the soliton as a function of its normalized amplitude δn /n.*

It has been observed that as the amplitude of the perturbation pulse is increased, the amplitude of half-cycles of the time-delayed pulses observed by the LPs increases. Simultaneously, the velocity of the propagation also increases. However, the width of the time-delayed pulses decreases as the perturbation amplitude increases, showing an opposite variation trend as compared to its amplitude.

The propagation velocity of the perturbation, obtained by the time-of-flight technique, has been found to be higher than the ion acoustic velocity ($C_s = \sqrt{K_B T_e/m_i}$)[36,68,69]. The Mach number ($u/C_s$) is plotted with respect to $\delta n/n$ in figure 9 and is shown in red line having diamond marker. Where $\delta n$ is the density perturbation and $n$ is the unperturbed plasma density. The measured Mach number variation with $\delta n/n$ matches very well with those calculated using the KdV equation given by Ikezi et.al[5,18], as shown by the blue line with star markers in the same figure. The spatial width of the soliton is measured experimentally using the standard technique[5,18–20,70,71], as discussed briefly below. First, the full width at half maximum of the positive part of the time-delayed pulses of the LP has been measured from the temporal evolution of LP data, as shown in figure 8a. This gives the temporal width of the structure. To obtain spatial width of the soliton, the measured temporal width is multiplied by the measured velocity of propagation. This gives the width of the soliton D, and following convention, the width is normalized by $\lambda_D$. The normalized width of the propagating structure is plotted as a function of the amplitude of the perturbation in figure 10 keeping all other plasma parameters constant at a cusp magnetic field value of $B_p = 0.6 kG$ (Magnet Current $I_{mag} = 80A$). It is observed from figure 9 that the amplitude of the propagating structures varies linearly with $u/C_s$ (Mach number).

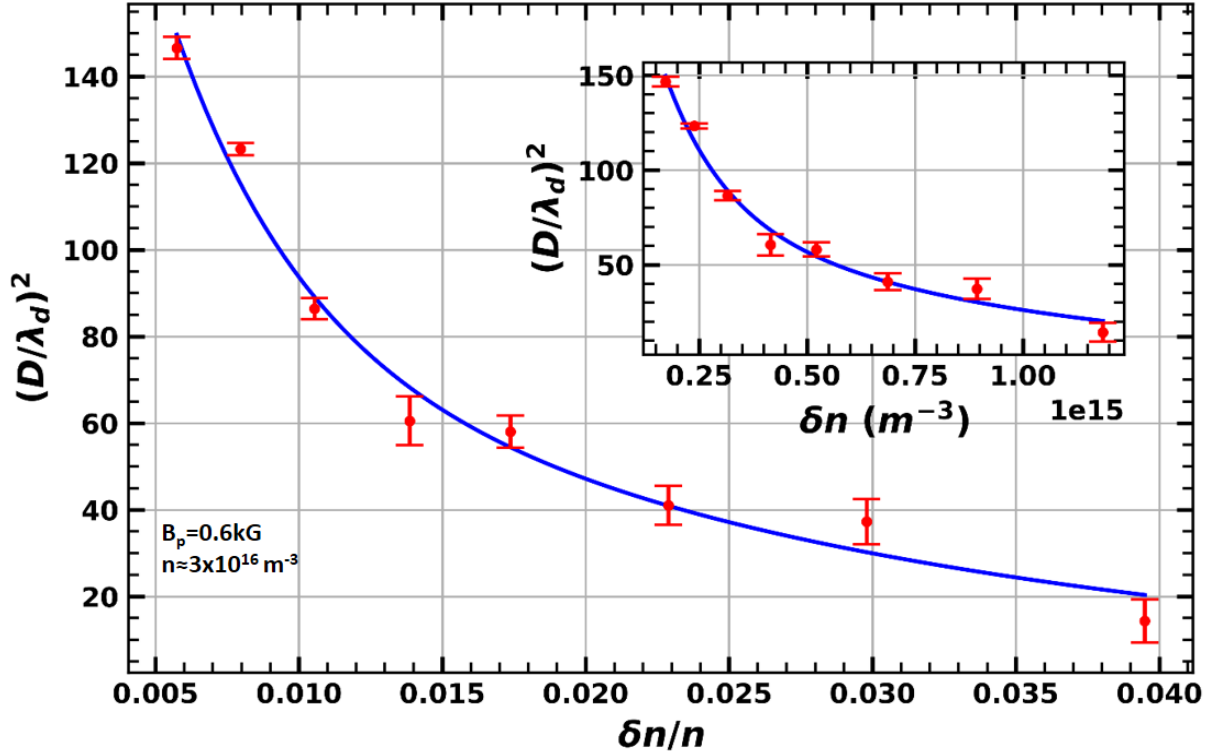

*Figure 10: The normalised width D of the soliton as a function of its normalised amplitude δn /n. The inset shows the variation of normalised width D with δn.*

Furthermore, it is observed from figure 10 that square of the width is inversely proportional to its amplitude. These observations are in accordance with the properties of small amplitude KdV [5,11,18,21,70,71] type of ion-acoustic soliton and hence indicate the excitation of solitons in our experiments.

In order to confirm further the propagating structures to be the solitons, two similar counter-propagating perturbations are generated in the MPD and made them interact with each other. It is quite well known that when two solitary waves collide, they overlap and pass through each other without losing their identity[5,11,24].

To excite the two counter-propagating solitons, another exciter disc of the same shape and size of the first one mentioned earlier, is placed on another side of the Langmuir probe set in the core-plasma. These probes and exciters are kept in a uniform field-free region and the location of grids and probes are shown in figure11. A perturbation amplitude of ~$20V_{pp}$ and frequency ~$90kH_z$ has been applied to both the exciters simultaneously. Figure 12 shows the interaction of two counter-propagating solitary waves. The bottom trace of the figure shows the two solitons, S1 (excited by grid 1) and S2 (excited by grid 2), are excited from the individual exciters and propagating towards each other. At time t = $t_0$, these solitons are picked by the probes 'probe-1' and 'probe-3', which are located adjacent and equidistant to 'grid-1' and 'grid-2' respectively.

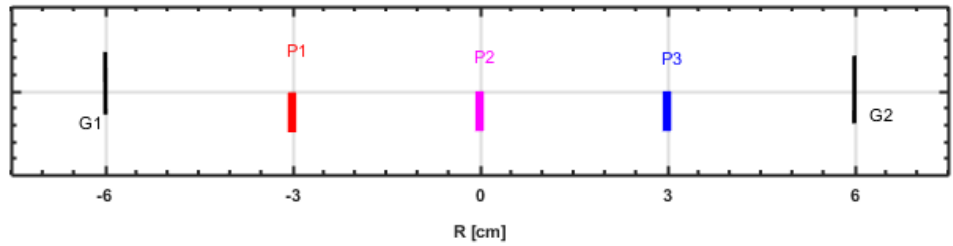

*Figure11: Schematic for locations of grids and probes for excitation of two counter propagating solitons*

As seen from the figure the solitons, S1 and S2 have almost same amplitude and velocity. As they move further, the solitons S1 and S2 interact at the center of the two exciters, where the 'probe-2' is located, merge into each other linearly and generate a single soliton, as shown in the mid trace of figure 12. After the interaction, they get separated and travel ahead with same velocity that they have prior to the interaction and without losing their identity. At this time (t = $t_0+46\mu s$), the soliton S1 excited by 'grid-1' is picked up by 'probe-3' whereas S2, excited by 'grid-2' is picked up by 'probe-1'. The above-mentioned observations further confirm that the propagating structures excited by two grids placed inside the MPD are solitons.

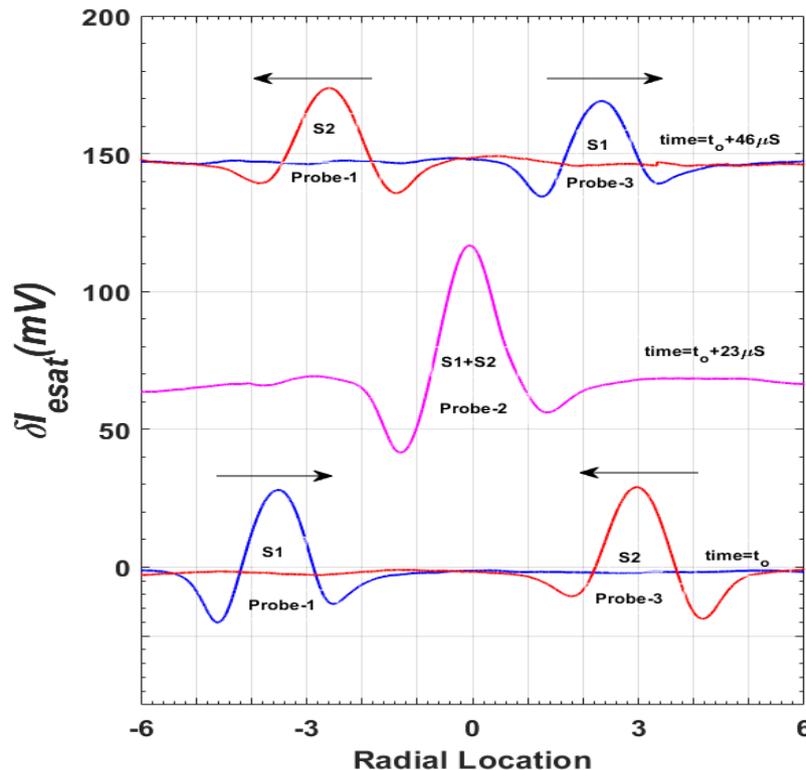

*Figure12: Interaction between two counter-propagating solitons. First Grid G-1 is located at R=-6cm and Second Grid G-2 is located at R=6cm. Probe-1,2 and 3 are 3cm,6cm and 9cm away from G-1. (For better visualization the traces are shifted vertically)*

# 4. Effect of Two-Electron Temperature on the Propagation of Ion Acoustic Soliton

After establishing the solitary nature of the propagating wave in MPD, the effect of two-temperature electron distribution on the propagation of IAS is studied by varying the ratios of the population of two-temperature electrons. As mentioned earlier, in MPD the electromagnets produce the cusp magnetic field, which gives freedom to change the pole cusp magnetic field strength by changing the applied currents to electromagnets. This change in cusp magnetic field strength also controls the population of cold and hot electrons in plasma confined by this magnetic field[32,41,43,44]. After exciting the IAS as described earlier, the pole cusp magnetic field has been varied by applying different magnitudes of currents to the electromagnets.

IAS is excited in the uniform field-free region where the ions are unmagnetized, and plasma is uniform and quiescent[32,37]. The cusp magnetic field configuration provides exceptional macroscopic plasma consistency due to U-shaped magnetic field curvature towards the confined plasma system in the center, and plasma is also stable to large-scale perturbation[44,72] and cusp field confines the maximum primary or high energetic electrons[41,73,44].

Figure 13 shows the variation of soliton amplitude and width with different pole cusp magnetic field values. It is observed that as the cusp magnetic field value is applied and increased initially, the soliton amplitude increases with a magnetic field. At $\sim 0.6\ kG$ ($I_{mag} = 80A$), the amplitude attains the maximum value. Increasing the cusp magnetic field further, the soliton amplitude starts decreasing gradually. During the initial increase of the cusp magnetic field where the amplitude of the soliton increases, the width has been observed to be decreasing, and beyond $\sim 0.6\ kG$ ($I_{mag} = 80\ A$), it starts increasing gradually. The observed inverse relation between the soliton amplitude and its width, as seen from figure 10, clearly demonstrates that the solitary nature of the triggered perturbation structure is sustained in the plasma at each applied cusp magnetic field value.

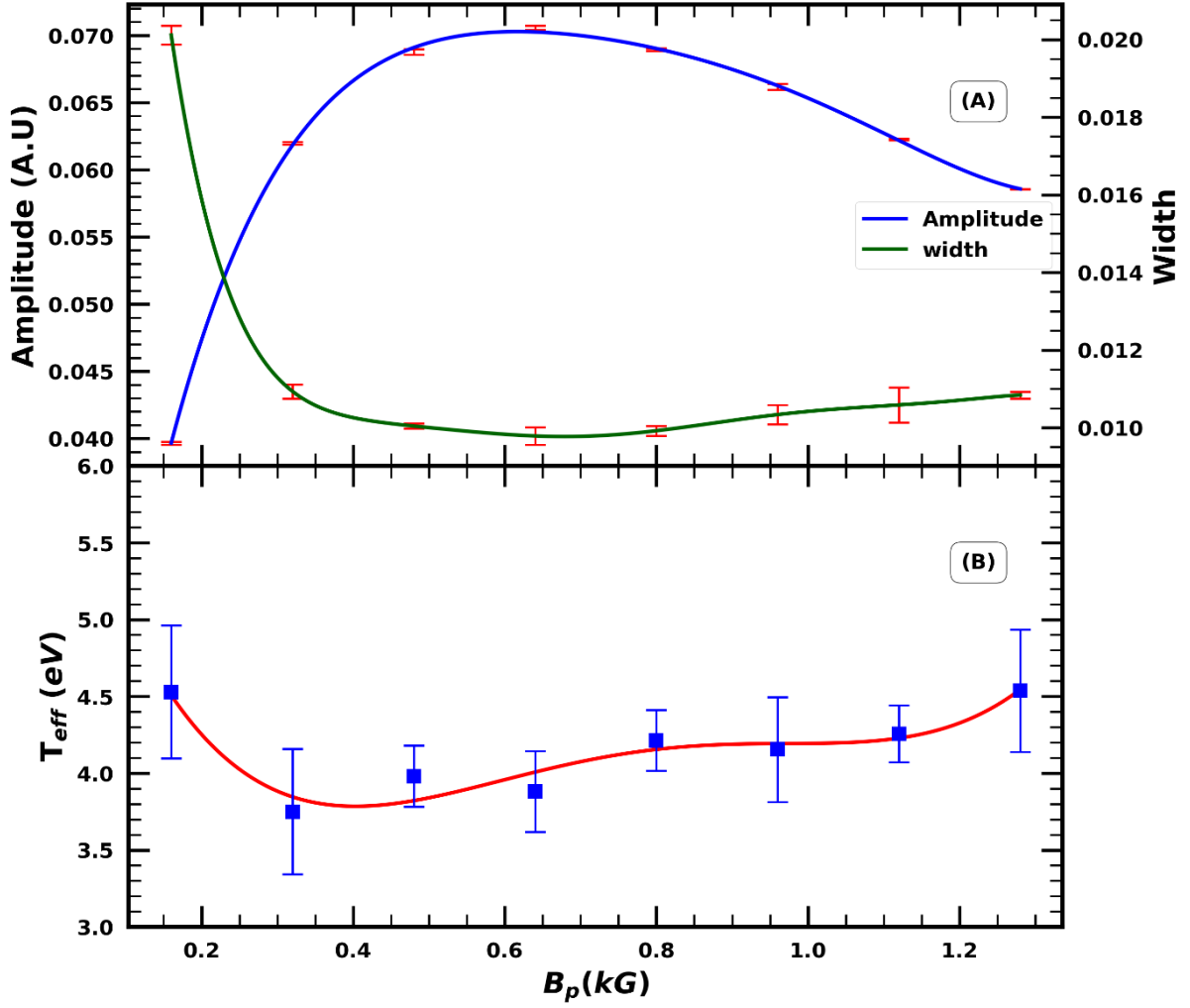

*Figure 13: (a) Variation of amplitude and width of soliton with increasing multi-pole cusp magnetic field strengths; (b) Variation of effective temperature with pole-cusp magnetic field strengths*

Variation of $T_{eff}$ with pole cusp magnetic field is also plotted in figure 13. It can be seen from the figure that the value of $T_{eff}$ initially decreases with the increase in the cusp magnetic field, reaching its minimum around $B_p = \sim 0.4 - 0.6 kG$ and then increase with the increase in the cusp magnetic field. The width of the soliton also varies in a similar fashion, whereas the amplitude of the soliton first increases and reaches to its maximum value at $B_p = \sim 0.4 - 0.6 kG$ before decreasing with an increase in the cusp magnetic field strength.

Soliton propagation being affected by $T_{eff}$ has not been studied much experimentally as varying $T_{eff}$ over a range in a single device, keeping the other parameters more or less constant, is quite difficult, and hence very few reports are available on the subject[60,74]. Taking advantage of MPD's unique feature of obtaining $T_{eff}$, the observations of variation of soliton amplitude and width with the $T_{eff}$ are very helpful in understanding the behaviour of soliton propagation in plasma having two temperature electrons in different fractions. Few theoretical analyses have reported the effect of $T_{eff}$ on the propagation of solitons. Goswami and Buti[60] have shown theoretically that as the $T_{eff}$ decreases, the amplitude of soliton increases. Though qualitatively, it agrees with the experimental results, and however, it does not explain the entire variation of soliton properties with the $T_{eff}$.

Lakhina[74] et.al, has shown through simulations that the amplitude of IAS gets modifies in presence of the high energetic electrons in the plasma. By solving the basic equations of the arbitrary amplitude solitons numerically, they have shown that the amplitude of the soliton is inversely proportional to the value of $T_{eff}$ i.e. as $T_{eff}$ increases (decreases), the soliton amplitude decreases (increases). Interestingly, similar behaviour of soliton propagation with the $T_{eff}$ has been observed in our experiments, substantiating the fact that the $T_{eff}$ indeed effect of the soliton propagation.

### **Summary & Discussion**

In this paper we report the excitation of the ion acoustic soliton in the MPD by applying a sinusoidal perturbation to a disk placed inside the field free region of the plasma. The propagating wave structures satisfy the relation between the amplitude, the Mach number, and width of the solitary wave and establishes the excitation of solitons in the experiments. By launching two counter-propagating perturbations and observing their overlapping and passing through each other without losing their identity ascertains the wave structures to be solitons. The maximum amplitude of the soliton generated in the present experiment is $B_p = 0.6 \, kG \, (I_{mag} = 80A)$. After thoroughly characterizing the existence of the solitons, the effect of two temperature electron distributions on the propagation of IAS has been explored. The effective temperature of electron ($T_{eff}$) has been varied by varying the population ratio and temperature of cold and hot components of electrons. It has been observed that in MPD pole cusp magnetic field value influences the propagation of IAS significantly. The amplitude of soliton has been found to be increasing with the field value up to $B_p = 0.6 \, kG$ after which it has been found to be decreasing with a further increase in the field values. The width of the soliton shows the opposite variation to its amplitude variation as a function of the cusp magnetic field. It has been observed that the evolution of solitons is sensitive to the effective temperature of plasma. Specifically, the amplitude and width of solitons vary significantly with $T_{eff}$. This observation quantitatively agrees with the theoretical study of the dependence of soliton amplitude with effective electron temperature in two-electron temperature plasmas.

## Acknowledgments

It is a pleasure to acknowledge Professor Abhijit Sen for fruitful discussions and encouragements. The authors are thankful to Dr. Pintu Bandopadhyay for the critical review of the manuscript. We are also thankful to of Dr. Sayak Bose and Mr. Rosh Roy for their helpful feedback and support. Z.S is indebted to University Grant Commission (UGC) and Ministry of Minority affairs, Government of India for their support under the Maulana Azad National Fellowship scheme award letter number 2016-17/MANF-2015-17-GUJ-67921.